# Remote sensing of angular scattering effect of aerosols in a North American megacity

Zhao-Cheng Zeng[1,2], Feng Xu[3], Vijay Natraj[3], Thomas J. Pongetti[3], Run-Lie Shia[1], Qiong Zhang[1], Stanley P. Sander[3] and Yuk L. Yung[1]

[1] Division of Geological and Planetary Sciences, California Institute of Technology, USA;

[2] JIFRESSE, University of California, Los Angeles, Los Angeles, USA；

[3] Jet Propulsion Laboratory, California Institute of Technology, USA.

Correspondence to: Z.-C. Zeng (zcz@gps.caltech.edu)

Address: 1200 E. California Blvd MC-150-21, Pasadena, CA 91125

**Abstract**

The angle-dependent scattering effect of aerosols in the atmosphere can be used to infer their compositions, which in turn is important to understand their impacts of human health and Earth's climate. The aerosol phase function, which characterizes the angular signature of scattering, has been continuously monitored from ground-based and space-borne observations. However, the range of scattering angles these instruments can sample is very limited. There is a dearth of research on the remote sensing of aerosol angular scattering effect at a city scale that analyzes diurnal variability and includes a wide range of scattering angles. Here, we quantify the aerosol angular scattering effect using measurements from a mountain-top remote sensing instrument: the California Laboratory for Atmospheric Remote Sensing Fourier Transform Spectrometer (CLARS-FTS). CLARS-FTS is located on top of the Mt. Wilson (1.67km above sea level) overlooking the Los Angeles (LA) megacity and receives reflected sunlight from targeted surface reflection points. The observational geometries of CLARS-FTS provide a wide range of scattering angles, from about 20° (forward) to about 140° (backward). The $O_2$ ratio, which is the ratio of retrieved $O_2$ Slant Column Density (SCD) to geometric $O_2$ SCD, quantifies the aerosol transmission with value of 1.0 represent aerosol-free and with value closer to 0.0 represents stronger aerosol loadings. The aerosol transmission quantified by the $O_2$ ratio from CLARS measurements provides an effective indicator of the aerosol scattering effect. Our results show that (1) CLARS-FTS measurements are highly sensitive to the angular scattering effect of aerosols in the LA urban atmosphere, due to the long light-path going through the boundary layer and wide range of observational angles; (2) the temporal patterns of the aerosol transmission over different surface reflection points are primarily determined by the diurnal changes of aerosol loadings and the scattering angles. Moreover, the difference in aerosol transmission between reflection points can be explained by the differences in their scattering geometries. The variability in angular scattering effects can be reproduced by a radiative transfer model using aerosol properties observed by AERONET-Caltech; and (3) correlation between measurements at different surface reflection points can be used to quantify the strength of the angular dependence of the aerosol phase function. Applying the correlation technique to the CLARS-FTS measurements, we found that, from 2011 to 2018, there is no significant trend in the aerosol composition in the the LA megacity. Overall, this study provides a practical observing strategy for quantifying the angular dependence of aerosol scattering in urban atmospheres that could potentially contribute towards monitoring urban aerosol composition in megacities.

**Research Highlights** (3-5 bullets; 85 characters including space)

- A mountain-top observatory for monitoring aerosols in megacity is introduced;
- The measurements are highly sensitive to aerosol scattering in the boundary layer;
- An aerosol transmission indicator based on oxygen retrieval is introduced;
- The aerosol scattering pattern can be explained by variation of scattering angle;
- No significant change in aerosol composition in LA was observed from 2011 to 2018.

# 1. Introduction

Aerosols in the atmosphere affect the Earth's energy balance directly by scattering and absorption of sunlight and indirectly through aerosol cloud interactions. It has been shown that the aerosols' direct and indirect effects are the two largest sources of uncertainties in quantifying anthropogenic radiative forcing (**Stocker et al., 2014**). Moreover, the adverse impact of aerosols on public health makes particulate matter air pollution the world's largest environmental health risk (**Heft-Neal et al., 2018**). A wide range of observation techniques, including space-based, airborne, and ground-based remote sensing, has been developed to provide constraints on the aerosol optical and microphysical properties. However, the composition of aerosols (varies by their sizes, shapes, and chemical characteristics), a key parameter for quantifying its impact on Earth's energy balance and human health, is under-constrained and therefore not well understood. One of the potential way to achieve aerosol speciation is using multi-angle measurements that provides additional constraints on the angular dependence of aerosol scattering (**Diner et al., 2018**). The aerosol phase function, which characterizes the angular signature of scattering, has been continuously monitored from ground-based and space-borne observations. The ground-based network of AERONET station has been monitoring the aerosol variability globally and retrieving their aerosol optical and microphysical properties, including phase function (**Kaufman et al., 1994; Dubovik et al., 2000**). Airborne measurements, e.g. AirMSPI (**Xu et al., 2016 and 2017**), have demonstrated the capability of multi-angle measurements for quantifying the aerosol properties over land and ocean. Multi-angle Imaging SpectroRadiometer (MISR; **Diner et al., 2005**) by NASA has been successful in retrieving aerosol microphysical properties associated with different anthropogenic and naturally occurring aerosol types (**Kahn et al., 2001**). The upcoming Multi-Angle Imager for Aerosols (MAIA) mission (**Diner et al., 2018**) will have significantly improved sensitivity to the composition of airborne particles with more spectral bands and advanced polarimetry. However, the range of scattering angles these instruments can sample is limited. Ground-based instruments mostly focus on forward scattering, whereas space-borne instruments are concentrated on backward scattering. There is a dearth of research on the remote sensing of aerosol angular scattering effect over a wider range of scattering angles (from backward to forward), especially observing at a city scales with diurnal capability.

In this study, we report multi-year (2011–2018) measurements of aerosol scattering from a mountain-top observing system: the California Laboratory for Atmospheric Remote Sensing-Fourier Transform Spectrometer (CLARS-FTS), which provides diurnal observations of aerosol scattering inside the Los Angeles (LA) megacity. LA, a sprawling urban area with 15 million people, has long been facing severe air pollution problems for the past half-century. The natural sources of aerosols include dust from the desert and sea salt from the ocean. The anthropogenic sources of aerosols in the atmosphere are mainly (1) soot directly from vehicle emissions and fire burning; (2) sulfates formed through sulfur dioxide emitted from power plants and other industrial facilities; (3) nitrates formed through nitrogen oxide emitted from vehicles and power plants; and (4) organic carbon formed through reactive organic gas emitted from vehicles, industrial facilities, forest fires, and biogenic sources (**EPA, 2003**).

CLARS-FTS instrument sits near the top (1.67 km a.s.l.) of Mt. Wilson and overlooks the LA megacity with its scanning capability. Measurements in the near infrared with high spectral resolution (0.06 cm$^{-1}$) are obtained continuously. The CLARS-FTS measurement system mimics geostationary satellite observations to some extent. From a fixed vantage points, it measures the reflected sunlight from surface and quantifies the atmospheric GreenHouse Gase (GHG) absorptions along the light path, which consists of optical paths from the sun to the land surface and from the land surface to CLARS-FTS. Because of its extended slant light path across the urban boundary layer, CLARS-FTS observing geometry makes the observations highly sensitive to anthropogenic emissions in the LA megacity (**Fu et al., 2013; Zhang et al., 2015; Wong et al., 2015; Wong et al., 2016; Zeng et al., 2017; Zeng et al. 2018; He et al., 2019**). In this study, we used measurements of oxygen absorption from CLARS-FTS to investigate the diurnal and seasonal variabilities of angular scattering due to aerosols in the LA basin. Since oxygen is a well-mixed gas with known concentration, the measurement is a proxy for the average length of light paths (direct and scattering) and indicator of the strength of aerosol scattering.

## 2. California Laboratory for Atmospheric Remote Sensing

### 2.1 CLARS-FTS

CLARS-FTS (**Figure 1**) is located at an altitude of 1.67 km near the top of Mt. Wilson, which is above the urban boundary layer (**Ware et al., 2016**). It overlooks 33 surface reflection points in the LA basin and offers high-resolution (~0.06 cm$^{-1}$) measurements of near-infrared spectra from 4000 to 8000 cm$^{-1}$ (**Fu et al., 2014; Zeng et al., 2018**). CLARS-FTS implements measurements using two observation modes: the Los Angeles Basin Surveys (LABS) mode and the Spectralon Viewing Observation (SVO) mode. In SVO mode, the FTS measures the reflected solar spectrum from an Spectralon® plate besides the FTS; In LABS mode, the FTS measures the reflected sunlight from the targeted regions on surface. The sunlight goes through a long light path within the LA megacity planetary boundary layer (PBL; **Figure 1(b)**) and undergoes strong absorption and scattering by air molecules and atmospheric particles in the basin. Such observation geometries make CLARS observations not only highly sensitive to the urban boundary layer atmospheric compositions, but also very susceptible to aerosol scattering effects due to strong aerosol loading in the basin. Slant column density (SCD) of trace gas is retrieved from CLARS-FTS measurements. SCD is the total number of absorbing trace gas molecules per unit area along the optical path. The details of CLARS-FTS observation system and its operational retrieval algorithms can be found in **Fu et al. (2014)**. In order to study the aerosol scattering effect, we retrieved the $O_2$ SCD at oxygen $^1\Delta$ band at 1.27 μm using the Gas Fitting tool (GFIT) algorithm, which has been used for several airborne infrared measurement experiments and the Total Carbon Column Observing Network (TCCON) network (**Toon et al., 1992; Wunch et al., 2011**). However, the aerosol scattering effect is not computed and taken into account in the GFIT retrieval algorithm. Therefore, bias in the retrieval of $O_2$ SCD primarily comes from the change of light path due to aerosol scattering effect; this bias can be used to investigate the aerosol optical properties (**Zhang et al., 2015; Zeng et al., 2018**). Surface reflectance of the surface reflection points can be calculated by the ratio of SVO-observed (incident solar spectrum) and LABS-observed (reflected sunlight) solar radiance under clear atmospheric conditions using continuum measurements (where gas absorption is negligible) around oxygen $^1\Delta$ band. The derived surface reflectance used in the radiative transfer (RT) model calculations are shown in **Appendix A1**.

For this study, we chose three surface reflection points located in the western San Gabriel Basin. They are West Pasadena (W-P), Santa Anita (S-A), and Santa Fe Dam (S-F), shown in **Figure 1(a)**. CLARS measurements over these three reflection points cover a wide range of aerosol scattering angles, as shown in **Section 4.1**. Moreover, the aerosols in this area originate from the LA downtown area since the sea breezes induced by land-sea thermal contrast bring the air pollution from the downtown westward to the San Gabriel Basin. The pollution in this region is especially severe in the afternoon, because the pollution may be trapped aloft when the mixed layer stabilizes (**Lu and Turco, 1994**). Observations of aerosol scattering from these three surface reflection points provide information on the aerosols from the LA megacity in general.

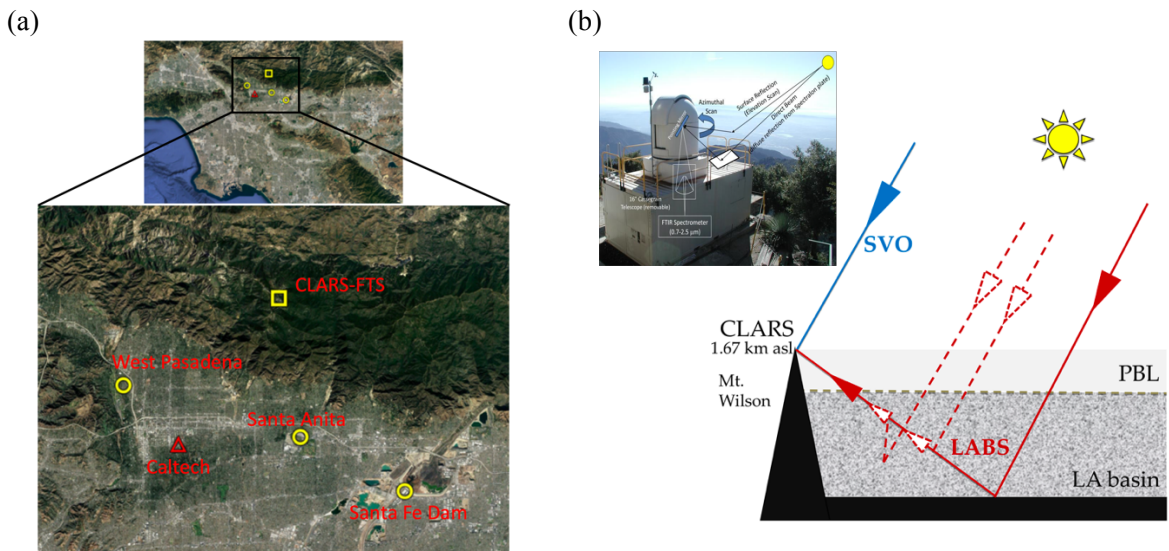

**Figure 1**. (a) CLARS-FTS near the top of Mt. Wilson and three of its surface reflection points in the LA basin: West Pasadena, Santa Anita, and Santa Fe Dam. The AERONET site at Caltech is also indicated. The background image is adopted from Google Maps; (b) Schematic diagram of CLARS-FTS observing geometries. CLARS-FTS implements measurements using two observation modes: the Los Angeles Basin Surveys (LABS) mode and the Spectralon Viewing Observation (SVO) mode. As an example, light path change (in dotted red) due to aerosol scattering along the optical path from the basin surface to the instrument is illustrated. The aerosol scattering includes contributions from single scattering and multiple scattering.

## 2.2 Aerosol transmission indicated by $O_2$ Ratio

In this study, we use the oxygen ($O_2$) SCD ratio (denoted by $O_2$ Ratio), which is the ratio of retrieved $O_2$ SCD (denoted by $O_{2,retrieved}$ SCD) to geometric $O_2$ SCD (denoted by $O_{2,geometric}$ SCD), as a proxy for light path change due to aerosol scattering:

$$O_2 \text{ ratio} = \frac{O_{2,retrieved} \text{ SCD}}{O_{2,geometric} \text{ SCD}} \quad [1]$$

The geometric $O_2$ SCD is derived from National Center for Environmental Prediction (NCEP) reanalysis data with known observing and solar geometries and a constant oxygen dry-air volume mixing ratio of 0.2095. We also assume hydrostatic equilibrium and no scattering or absorption along the optical path (**Zhang et al., 2015**). In a non-scattering atmosphere, $O_2$ SCD retrievals from CLARS-FTS should have the same value as the geometric $O_2$ SCD. The Rayleigh scattering contribution is negligible since we are using the measurements at near infrared band of 1.27um. In a scattering atmosphere with aerosols, the change of optical light path due to aerosol scattering effect makes the retrieved $O_2$ SCD deviate from the calculated geometric $O_2$ SCD. Therefore, the deviation of $O_2$ ratio from unity provides a proxy for the extent of aerosol scattering over the basin, when the deviation is larger than the retrieval uncertainty, which is 0.5% for LABS measurements (**Fu et al., 2014; Zeng et al., 2018**). This approach of utilizing $O_2$ retrieval as proxy for scattering effect is equivalent to that used for GOSAT (**Yokota et al., 2009**) and OCO-2 (**Crisp et al., 2008**) retrievals, $O_2$ A-band observations are compared with reanalysis data to discriminate the light path-induced changes from changes in actual trace gases (**O'Dell et al., 2012; Taylor et al., 2016**). The $O_2$ $^1\Delta$ absorption band at 1.27 μm used by CLARS-FTS for retrieving $O_2$ SCD is shown in **Figure 2**. In summary, the $O_2$ ratio value effectively quantifies the strength of aerosol scattering and can be defined as aerosol transmission. The rule of thumb for using this aerosol transmission as indicated by $O_2$ ratio to quantify aerosol scattering effect is: the lower of the aerosol transmission indicator, the stronger of the aerosol scattering. If aerosol transmission indicator is 1.0, then there is no aerosol scattering effect.

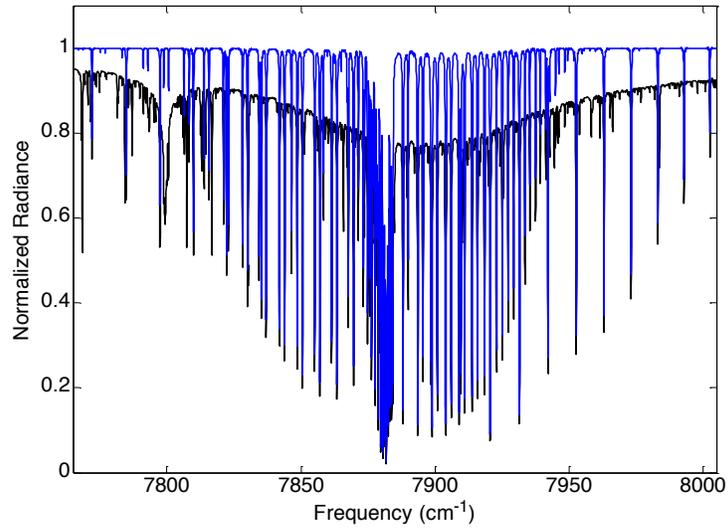

**Figure 2**. An example of the normalized radiances in the oxygen $^1\Delta$ band at 1.27 μm, centered at 7885 cm$^{-1}$ (1.27 μm) with width of 240 cm$^{-1}$, selected for retrieving O$_2$ SCD from CLARS-FTS measurements. The black lines indicate CLARS-FTS measurements, including contributions from all trace gases and solar lines at a solar zenith angle of 41.45°. The blue lines show the estimated contribution from O$_2$ line by line absorptions (O$_2$ collision induced absorption is not included here) to the absorption spectra calculated by the 2S-ESS RT model (**Section 3**).

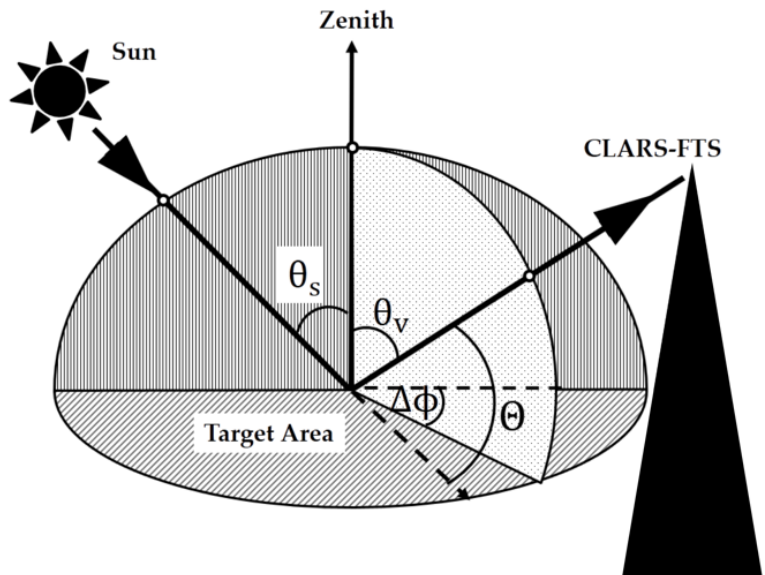

**Figure 3.** Illustration of the observing geometries of CLARS-FTS, including the scattering angle ($\Theta$), solar zenith angle ($\theta_s$), viewing zenith angle ($\theta_v$) and relative azimuth angle ($\Delta\phi$). The scattering angle can be calculated from the zenith angles and the relative azimuth angle using **Equation (2)**.

## 2.3 Aerosol Scattering Angle and phase function

The scattering angle is defined as the angle between the incident and scattered light beams, as shown in **Figure 3**. From spherical geometry (**Liou, 2002**), the scattering angle can be derived from the incoming and outgoing directions:

$$\cos(\Theta) = \mu\mu' + (1-\mu')^{1/2}(1-\mu'^2)^{1/2}\cos(\phi'-\phi) \qquad [2]$$

where $\mu$ and $\mu'$ are the cosines of solar and viewing zenith angle, respectively, and $\phi$ and $\phi'$ are the solar and viewing azimuth angles, respectively. The aerosol scattering phase function defines the angular distribution of aerosol scattering energy in terms of scattering angle. **Figure 4** shows the scattering phase functions following the Henyey-Greenstein approximation for various values of the asymmetric factor *g*. The Henyey-Greenstein phase function (**Henyey and Greenstein, 1941**) can be analytically expressed as:

$$P_{HG}(\Theta) = \frac{1}{4\pi}\frac{1-g^2}{(1+g^2-2g\cos(\Theta))^{3/2}} \qquad [3]$$

g=0.0 corresponds to isotropic scattering, where the scattering energy is the same in all directions. As the peak of the phase function sharpens in the forward direction, g increases. In **Figure 4**, g=0.67 is the averaged asymmetric parameter at 1020 nm derived from the AERONET retrievals at Caltech (see **Appendix A3**). As a comparison, the phase functions for the five types of aerosols obtained from the MERRA aerosol reanalysis data (**Rienecker et al., 2011**) and calculated using the GOCART model (**Colarco et al., 2010**) are also shown. It clearly shows that the averaged phase function from AERONET ranges between phase function of aerosols from natural sources (dust and sea salt) and those from mostly anthropogenic sources (black carbon, organic carbon and sulfate).

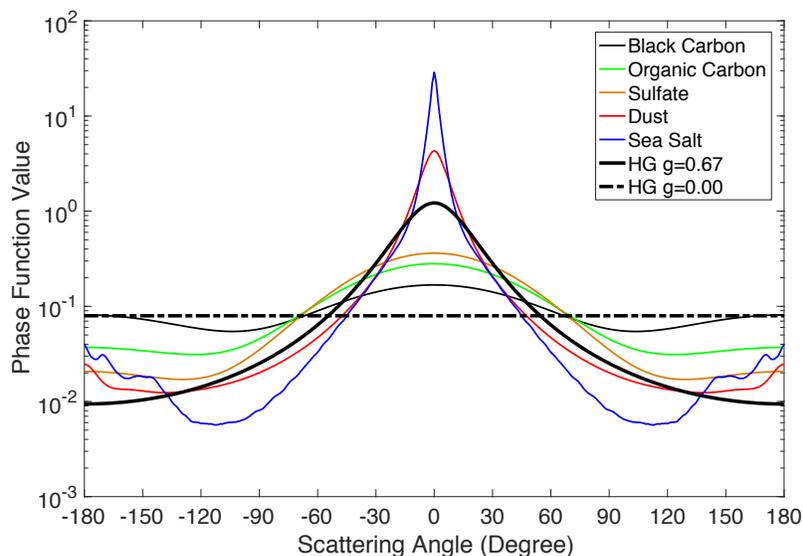

**Figure 4**. The aerosol scattering phase function for five different aerosol types, including black carbon, durst, organic carbon, sea salt, and sulfate, and two phase functions following the Henyey-Greenstein approximation with asymmetry parameter (g) values of 0.67 and 0.0, respectively. The value g of 0.67 is the averaged asymmetry parameter in Pasadena inferred from AEROENT-Caltech. The five types of aerosols are obtained from the MERRA aerosol reanalysis data (**Rienecker et al., 2011**) and calculated using the GOCART model (**Colarco et al., 2010**).

## 3. Radiative transfer model for aerosol scattering in LA megacity

The Two-Stream-Exact-Single-Scattering (2S-ESS) RT model (**Spurr and Natraj, 2011**) is used to simulate the observations and quantify the effect of aerosol scattering on the changes in $O_2$ ratio retrieved from CLARS-FTS. It has been widely used for remote sensing of trace gases in previous studies (**Xi et al., 2015; Zhang et al., 2015; Zhang et al., 2016; Zeng et al., 2017; Zeng et al., 2018**). The settings of the RT model follow those used by **Zhang et al. (2015)** and **Zeng et al. (2018)**. Basically, the *a priori* profile of atmospheric composition is obtained from NCEP-NCAR reanalysis data (**Kalnay et al., 1996**); the absorption coefficients for the oxygen molecules and the optical depths for all layers are computed using the GFIT program (**Toon et al., 1992**); the surface reflection is derived from CLARS-FTS measurements and shown in **Appendix A1**; Rayleigh scattering is calculated and incorporated in the model; the CLARS observing viewing zenith angle, the solar zenith angle, and the relative azimuth angle in between are included; the phase function of aerosol scattering in the RT model is assumed to be the Henyey-Greenstein approximation (**Henyey and Greenstein, 1941**). Below the CLARS altitude, the atmosphere is divided into 5 layers. Aerosols in these layers are assumed to be well mixed. The aerosol optical properties, including single scattering albedo and asymmetry parameter from phase function, are adopted from AERONET measurements at Caltech; the generated radiance from the RT model is convolved with instrument line shape (ILS) from CLARS-FTS (**Fu et al., 2014**). The spectral resolution of the radiance is set to be the same with CLARS measurement (0.06 cm$^{-1}$). We assume the signal-to-noise ratio (SNR) to be a constant of 300. Gaussian white noise is then added to the simulated spectra. The aerosol optical depth (AOD) measurements from AERONET-Caltech cover from 340 nm to 1020 nm; however, the $O_2$ $^1\Delta$ absorption bands (7885 cm$^{-1}$, ~1270 nm) used in this study are outside that range. The Ångström exponent law (**Seinfeld and Pandis, 2006**) is used to extrapolate the AOD data at 1270 nm. The AOD data are added in the PBL in the RT model assuming even vertical and horizontal distribution.

To quantitatively evaluate the effect of aerosol scattering on the retrieval of $O_2$ SCD from CLARS-FTS, we estimate the retrieval bias caused by aerosol scattering by a two-step process. First, synthetic spectral radiance data are generated by the 2S-ESS RT model with observed AOD from AERONET-Caltech; second, $O_2$ SCD is retrieved by fitting the synthetic spectra using the RT model based on Bayesian inversion theory (**Rodgers, 2000**). In the retrieval, the used RT model has the same configurations but with AOD set to zero and not retrieved. As shown in Zeng et al. (2018), this two-step process reproduces the $O_2$ SCDs by CLARS and approximately quantifies the effect of neglecting aerosol scattering on the retrieved. The non-linear Levenberg-Marquardt algorithm in the Bayesian inversion is employed for fitting the spectra (**Rodgers, 2000**). A scaling factor, *viz.*, the ratio of retrieved $O_2$ SCD to the geometric $O_2$ SCD derived from NCEP reanalysis data, is the state vector element to be retrieved using the Bayesian inversion approach. This scaling factor is equivalent to the $O_2$ ratio.

## 4. Results

### 4.1 Diurnal variability of aerosol scattering angle

The viewing zenith angles for the W-P, S-A, and S-F surface reflection points are 83.13°, 80.48°, and 84.07°, respectively. For these three surface reflection points, the reflected light goes through 8.36, 6.04, and 9.68 times the air mass, respectively, of nadir mode geometry in the PBL where most urban pollutants are trapped. Further, the observation geometries span a wide range of aerosol scattering angles. As illustrated in **Figure 5**, the S-F measurements are in the forward scattering in the morning, moving to backward scattering in the afternoon. The reverse is the case for W-P. A detailed quantitative description of diurnal scattering angle changes are in **Figure 6**. For S-F, the scattering angle increases from 30° (forward scattering) in the morning to 120° (backward scattering) in the afternoon. Conversely, for W-P, the scattering angle decreases from 140° (backward scattering) in the morning to 20° (forward scattering) in the afternoon. Interestingly, the scattering angle change of S-A is much smaller compared to W-P and S-F. For all seasons, the change for S-A is less than about 20°. Moreover, from **Figure 6**, the scattering angles have seasonal dependence due to the change in the geometries of the incoming solar beam. The scattering angles in summer are generally larger than those in winter. Considering such distinct diurnal variability of scattering angles at these surface reflection points and the long reflected light path within the PBL, CLARS-FTS measurements are highly sensitive to the scattering effects due to urban aerosols.

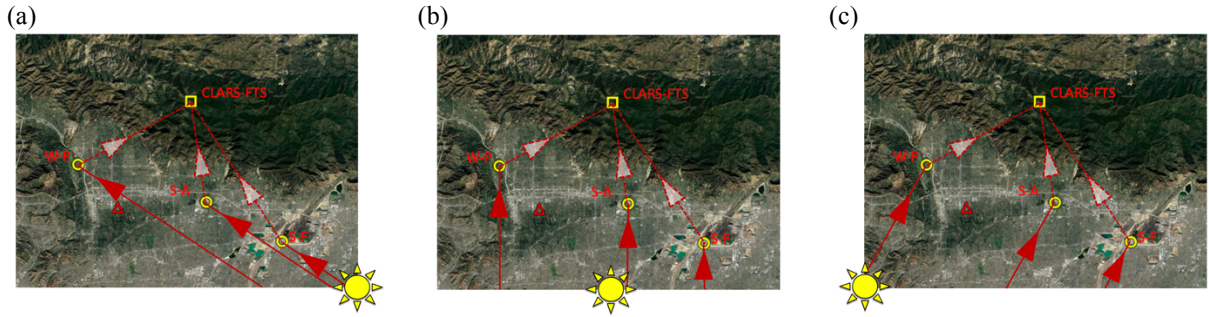

**Figure 5**. Illustration of incident (solid red) and reflected (dashed red) sunlight at the three surface reflection points (W-P for West Pasadena, S-A for Santa Anita, and S-F for Santa Fe Dam) for three different times of the day: (a) morning, (b) noon, and (c) afternoon. The light paths are the projections on the surface. The actually light paths in 3D space are not shown.

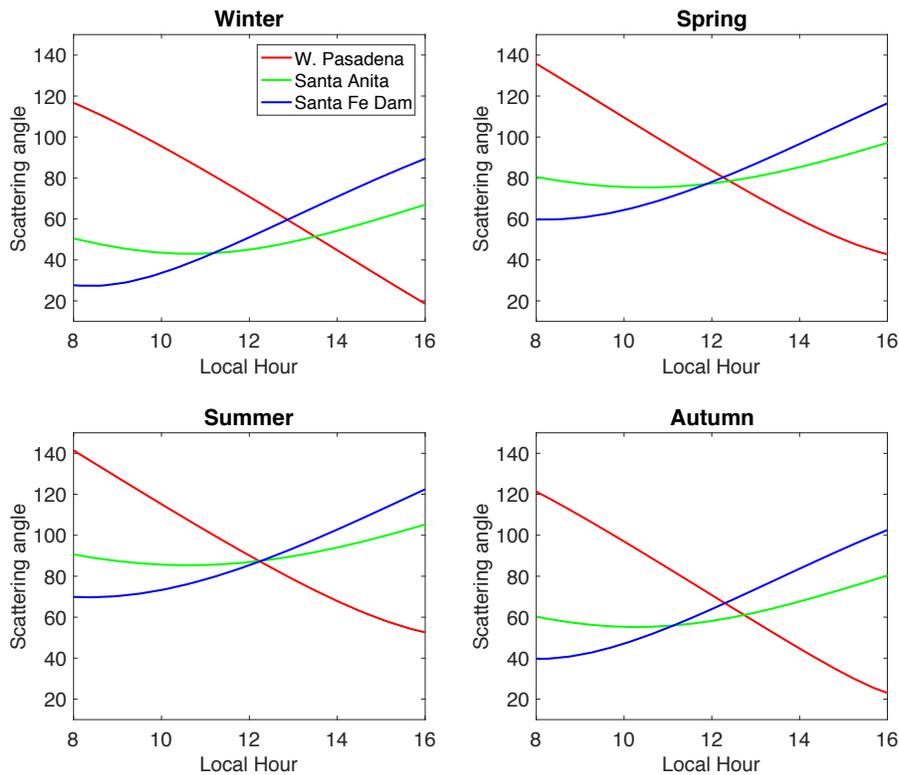

**Figure 6.** Diurnal variation of aerosol scattering angle, calculated using **Equation (3)**, from 8.0 h to 16.0 h local time over the three surface reflection points of West Pasadena, Santa Anita, and Santa Fe Dam for four different seasons: Winter, Spring, Summer, and Autumn, represented by the observing geometries on Jan 15, April 15, July 15, and October 15, respectively.

## 4.2 Diurnal variability of aerosol scattering effect

Since all the reflection points are located very close to the AERONET-Caltech site (the distances are 5.0, 7.7 and 14.9 km, respectively for W-P, S-A, and S-F), it is reasonable to assume that the vertical aerosol loadings at these three locations are almost the same. Therefore, the diurnal difference in aerosol scattering effects can be primarily attributed to the differences in scattering angle, given that the surface reflectance can be estimated with high accuracy (**Appendix A1**). **Figure 7** shows the $O_2$ ratio retrievals from CLARS-FTS for W-P, S-A, and S-F. Even though the observed AOD keeps increasing from the morning to the afternoon (**Appendix A2**), the changes in $O_2$ ratio for S-A are relatively small, and remain stable in the afternoon. This is because the change in scattering angle for S-A is smaller compared to W-P and S-F. The $O_2$ ratio for W-P, however, decreases from the morning to the afternoon, corresponding to enhanced aerosol scattering effects from two effects: increase in AOD and change from backward to forward scattering. On the other hand, the $O_2$ ratio for S-F has a small increase from the morning to the afternoon, indicating a weakening scattering effect. This is due to the sharp change in scattering angle from forward to backward scattering (**Figure 6**), partially offset by the increasing AOD. Such diurnal patterns are very coherent over all the measurements (over eight years) at the surface reflection points, as shown in **Figure 8**. S-F has larger variability compared to W-P and S-A because of the competition between AOD and scattering angle effects. The increasing AOD and increasing scattering angle has opposite effects on the S-F $O_2$ ratio. The same conclusions can be drawn when $O_2$ ratios are plotted against the scattering angle, as shown in **Appendix A4**.

The simulation results from the 2S-ESS RT model are shown as solid and dashed lines in **Figure 7** and as red lines in **Figure 8**. Overall, the RT model qualitatively reproduced the diurnal variation of $O_2$ ratios for all surface reflection points. The small differences between the measurements and the simulations may be due to the simplified assumption of H-G phase function and the usage of identical vertical aerosol loadings at all reflection points. To further investigate the effect of angular scattering dependence, a control run as shown in **Figure 7(b)** was conducted. We kept all the same settings as for **Figure 7(a),** except that the asymmetry parameter of the aerosol phase function was changed to zero, i.e., scattering was assumed to be isotropic. Therefore, any changes in the simulation results can be attributed to the angular scattering effect. From **Figure 7(b)**, we can see that the distinctive diurnal patterns of W-P, S-A, and S-F disappear. All changes in $O_2$ ratio primarily follow the change in AOD. The results from this experiment suggests that the variation in scattering effects between W-P, S-A, and S-F CLARS-FTS measurements is dominantly driven by the angular distribution of aerosol scattering in the LA basin.

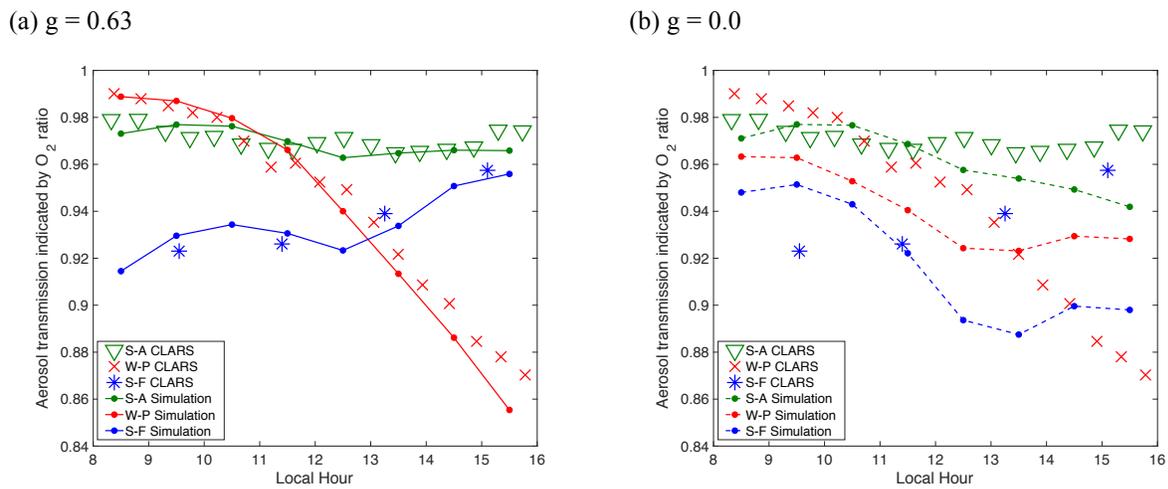

**Figure 7**. Comparison of $O_2$ SCD ratio, which is the ratio of retrieved $O_2$ SCD to geometric $O_2$ SCD, between measurements from CLARS and simulations from the 2S-ESS RT model over the West Pasadena (W-P), Santa Anita Park (S-A), and Santa Fe Dam (S-F) surface reflection points. The inputs for the 2S-ESS RT model are described in the **Section 3**. (a) the model simulations with asymmetry parameter (g=0.67) derived from the average of AERONET-Caltech data from 2011 to 2018; (b) control experiment (g=0.0) of model simulations assuming no angle-dependent aerosol scattering effect.

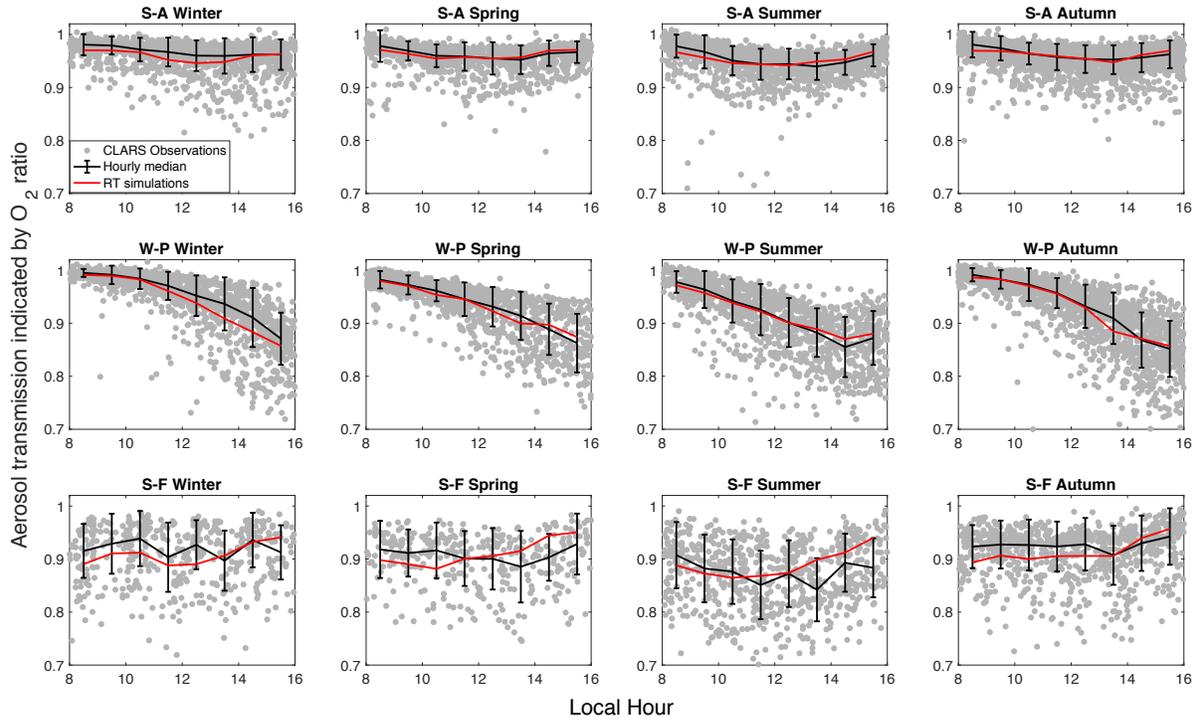

**Figure 8**. Multiyear average of $O_2$ ratio for the three surface reflection points, Santa Anita (S-A) on the top, West Pasadena (W-P) in the middle, and Santa Fe Dam (S-F) at the bottom from CLARS-FTS data for all seasons (original measurements in grey and hourly mean in black). The corresponding simulation estimates using the 2S-ESS RT model with averaged solar geometry, aerosol optical properties, and surface reflectance (see **Appendix A1**). All the data from 2011 to 2018 are used. **Appendix A4** shows the same plots but with the scattering angles as the x-axis.

## 4.3. Inter-annual variability of aerosol scatterings

The interannual variability of the aerosol phase function may indicate changes in aerosol compositions. For example, if there is an increase in fine mode particles from anthropogenic sources (e.g., black carbon, organic carbon, and sulfate) increases, the asymmetry parameter will become smaller since the angular dependence becomes weaker (as indicated by the phase function change in **Figure 4**). On the other hand, if coarse mode fraction from natural sources (e.g., dust or sea salt) increases, the angular dependence will become stronger. Here, we develop a correlation technique between measurements at surface reflection points to quantify the strength of the angular dependence of the aerosol phase function. This technique investigates the correlation between the $O_2$ ratios from W-P and S-A in the late afternoon (14-16h in this study) when they have very different scattering angles. RT model simulations are shown in **Figure 9(a)**; we can see that the $O_2$ ratios are highly correlated between W-P and S-A. The slopes between them dependes on the asymmetriy parameter. To the first order, if the asymmetry parameter is higher, the slope between W-P and S-A will be higher. This correlation technique provides a way to quantify the anisotropy of the phase function.

We applied this correlation technique to the CLARS-FTS measurements as shown in **Figure 9(b)**. A strong correlation between W-P and S-A can be seen here. The real measurements are much more noisy because the true aerosol properties have higher variability than those assumed in the RT model simulations, which can affect the retrievals. It is evident that the CLARS-FTS measurements are highly mixed for different years, such that one can hardly disentangle the measurements by years. A linear regression is applied to fit the data for different years with the regression line forced to cross the [1,1] point. The hypothesis here is that, if there is an interannual trend in the asymmetry parameter, there will be a trend in the regression slopes between W-P and S-A $O_2$ ratios. **Figure 9(c)** shows the time series of the regression slopes and the comparison with the annually averaged asymmetry parameter from AERONET-Caltech. We can see that there is no significant trend from 2011 to 2018, inferred from both CLARS-FTS and AERONET data. Overall, this study provides a practical observing strategy for quantifying the angular dependence of aerosol scattering in urban atmosphere that could potentially contribute toward monitoring urban aerosol types in megacities.

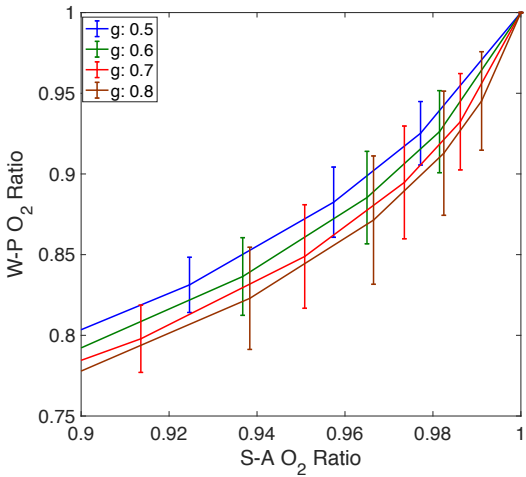
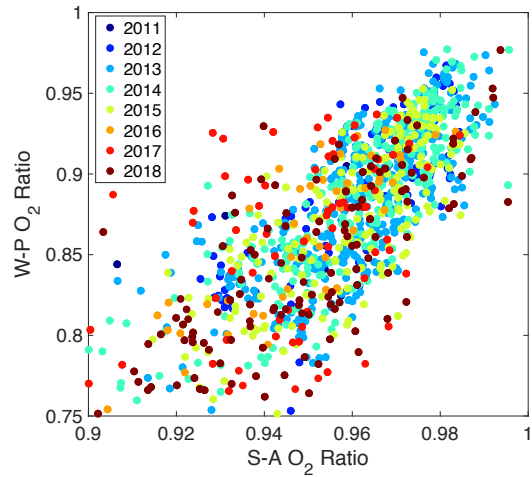
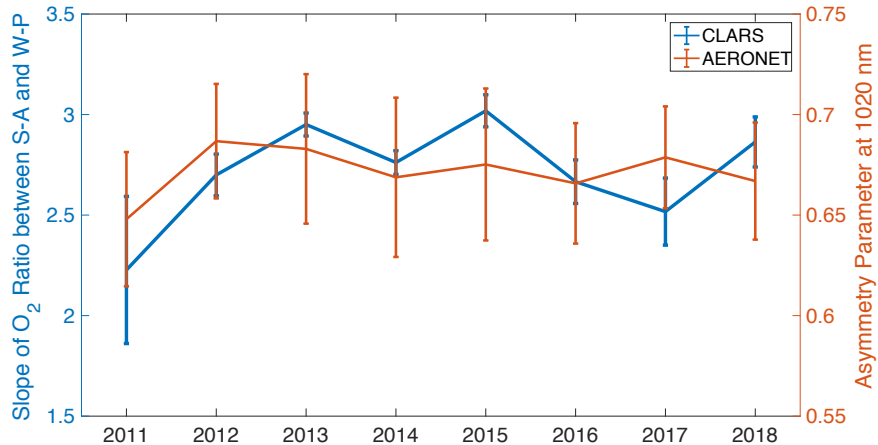

**Figure 9.** (a) Correlation of $O_2$ ratios between West Pasadena (W-P) and Santa Anita (S-A) surface reflection points for different asymmetry parameters (g) from 0.5 to 0.8 using the 2S-ESS RT model. The observing and solar geometries and all other related RT model inputs between 14h and 16h for all seasons are used for the RT simulation. The mean values are plotted and their standard deviations are shown as the error bars; (b) Scatter plot of $O_2$ ratio between W-P and S-A between 14h to 16h from CLARS-FTS retrievals from 2011 to 2018; (c) Regression slopes of $O_2$ ratio between W-P and S-A and comparison with annually averaged asymmetry parameter from AERONET-Caltech. The regression slopes are derived from applying a linear regression to fit the data in (b) for different years with the regression line forced to cross the [1,1] point.

## 5. Discussion

### 5.1 Use of oxygen ratio as indicator of aerosol scattering

The well-known mixing ratio and absorption spectroscopy of $O_2$ in the atmosphere make it a perfect indicator of aerosol scattering effects (**Yamamoto and Wark, 1961; Zeng et al., 2018**). The $O_2$ ratio indicator for aerosol scattering has been widely used in analyzing remote sensing satellite data (see, e.g., **Sanders et al., 2015; Taylor et al., 2016**). In this study, we use the ratio of retrieved to geometric $O_2$ SCDs instead of absolute radiance to study the aerosol scattering effects, for two main reasons: (1) $O_2$ ratio, which quantifies the aerosol transmission, has higher sensitivity to the ground level aerosols since the air density is higher and there are more $O_2$ molecules per unit volume. Therefore, the $O_2$ ratio as defined has advantages for detecting changes in aerosol optical properties close to the source of anthropogenic emissions; (2) understanding aerosol scattering effects on trace gas retrievals will help understand the impacts of light path uncertainty on the retrieval of trace gases from space, such as for the recently launched OCO-3 mission (**Eldering et al., 2019**). However, if the aerosols are optically thick, the $O_2$ ratio may lose sensitivity since the $O_2$ absorption will be saturated.

### 5.2 Application of CLARS aerosol observation systems to other cities

This measurement system can be used for many other megacities with vantage points above high buildings or mountains. Even if the vantage points are not above the boundary layer, the measurement system can be optimized for measurements of aerosol optical properties such as phase function and single scattering albedo, which do not require the instrument to sit above the PBL. This is different from measuring total aerosol loading and trace gas columnes, which do have this requirement in order to get the best observation. The surface reflection points should be carefully selected to have surfaces with homogeneous characteristics. The universal measurement system for all cities will be similar to the current open-path remote sensing instrument (**Griffith et al., 2018; Byrne et al., 2019**) above a high tower or building for trace gases studies. Such open-path instruments can be adjusted to monitor the aerosol optical properties at a city scale.

## 6. Conclusions

Quantifying the angular scattering effect of urban aerosols at a city scale has been challenging due to a lack of appropriate observing systems. We introduced the CLARS-FTS observing system, which overlooks the LA megacity mimicking a geostationary satellite. The angular scattering effects of aerosols are quantified by careful selection of surface reflection points that allow the observatory to measure the scattered light at different scattering angles. Using measurements from 2011 to 2018, we found that (1) the observation geometries from a mountain-top FTS over the LA megacity make it feasible to examine the aerosol scattering effect from forward to backward scattering. The long light path through the PBL makes the observations highly sensitive to aerosol scattering within the boundary layer; (2) The diurnal variability of aerosol scattering show distinct patterns for different surface reflection points, which are controlled by the total aerosol loading and changes in the scattering angle; (3) The changes in the angular scattering effects between the surface reflection points can be used to infer the changes in aerosol composition. Analysis of CLARS measurements from 2011 to 2018 showed no significant changes in aerosol composition during this timeframe in the LA megacity. Results from this study have implications for future trace gas observing satellites with observation geometry different from nadir mode, such as the target mode of OCO-3. The accuracy of the retrievals will heavily rely on the characterization of aerosol scattering phase function among other aerosol optical properties. As part of a future effort, we will apply the developed method to the measurements at all 33 surface reflection points (**Wong et al., 2015**) distributed all over the LA basin in order to understand the aerosol angular scattering effect at larger scales with distinct sources of anthropogenic emissions.

**Acknowledgement**

The CLARS project receives support from the California Air Resources Board and the NIST GHG and Climate Science Program. V. N. acknowledges support from the NASA Earth Science US Participating Investigator program (solicitation NNH16ZDA001N-ESUSPI). F. X. acknowledges support from the NASA Remote Sensing Theory program under grant 14-RST14-0100. We are also thankful for the support from the Jet Propulsion Laboratory Research and Technology Development Program. AERONET data for the Caltech site are available from https://aeronet.gsfc.nasa.gov/new_web/photo_db_v3/CalTech.html. We also thank Jochen Stutz from UCLA and his staff for their effort in establishing and maintaining the AERONET Caltech site. CLARS-FTS data are available from the authors upon request, and part of the data are available from the NASA Megacities Project at https://megacities.jpl.nasa.gov.

**Declarations of interest:**

none

# APPENDIX

## A1. Surface reflectance at West Pasadena, Santa Anita, and Santa Fe Dam.

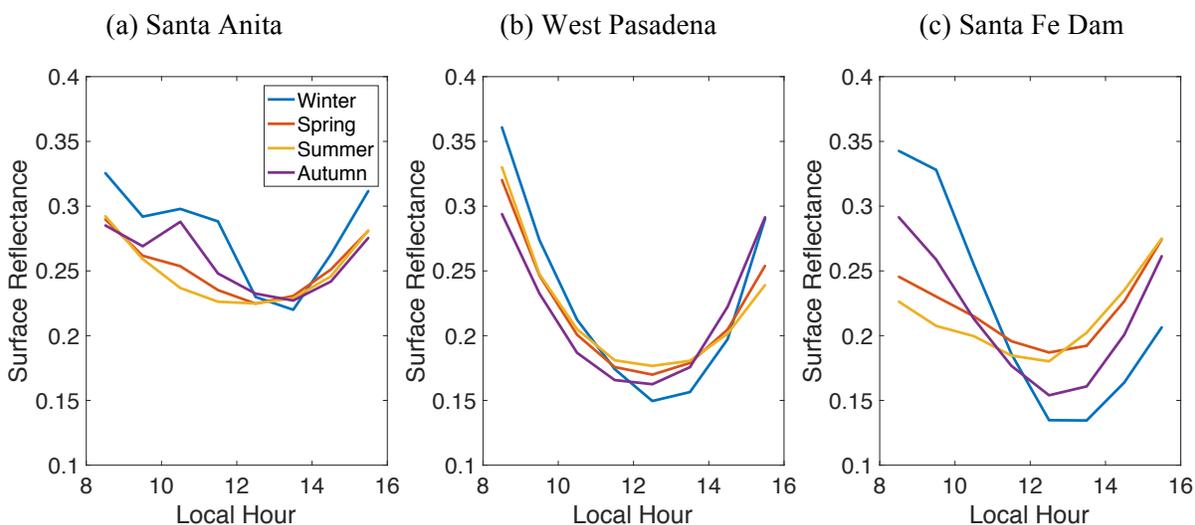

**Figure A1**. Averaged hourly surface reflectance at (a) West Pasadena, (b) Santa Anita, and (c) Santa Fe Dam for different seasons. Surface reflectance of the surface reflection points can be calculated by the ratio of SVO-observed (incident solar spectrum) and LABS-observed (reflected sunlight) solar radiance under clear atmospheric conditions using continuum measurements (where gas absorption is negligible) around oxygen $^1\Delta$ band. Clear days are chosen to be days with minimum reflected radiance for each month based on all measurements from 2011 to 2018.

## A2. The aerosol optical depth from AERONET

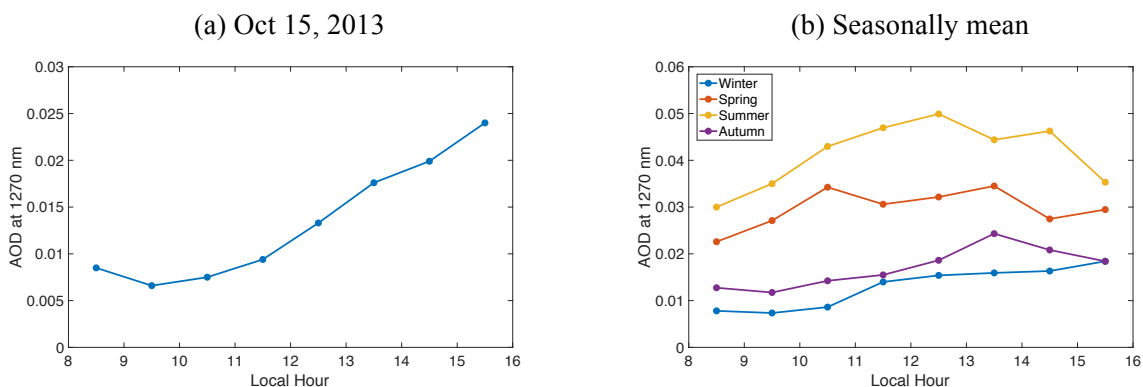

**Figure A2.** The diurnal AOD from AERONET-Caltech for (a) Oct 15, 2013; and (b) four seasons. The AOD are sampled according to the observing time of the available CLARS measurements after filtering. Since CLARS observations have filtered out a portion of the measurements on heavily hazy days, therefore, some of the extreme high AOD has also been excluded before making these hourly means.

## A3. Aerosol phase function from AERONET-Caltech

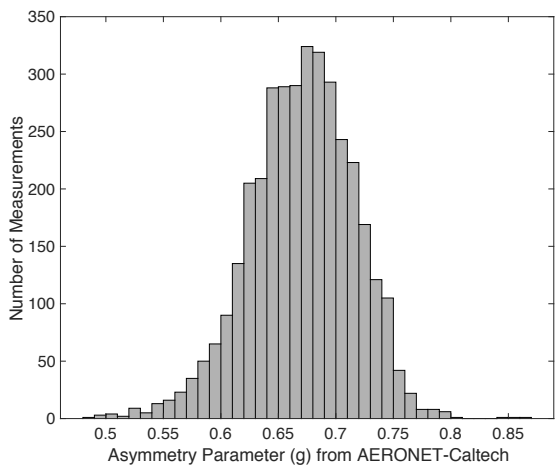

**Figure A3**. Histogram of asymmetric parameter at 1020 nm from AERONET-Caltech from 2011 to 2018.

## A4. Variabilities of aerosol transmission with scattering angles

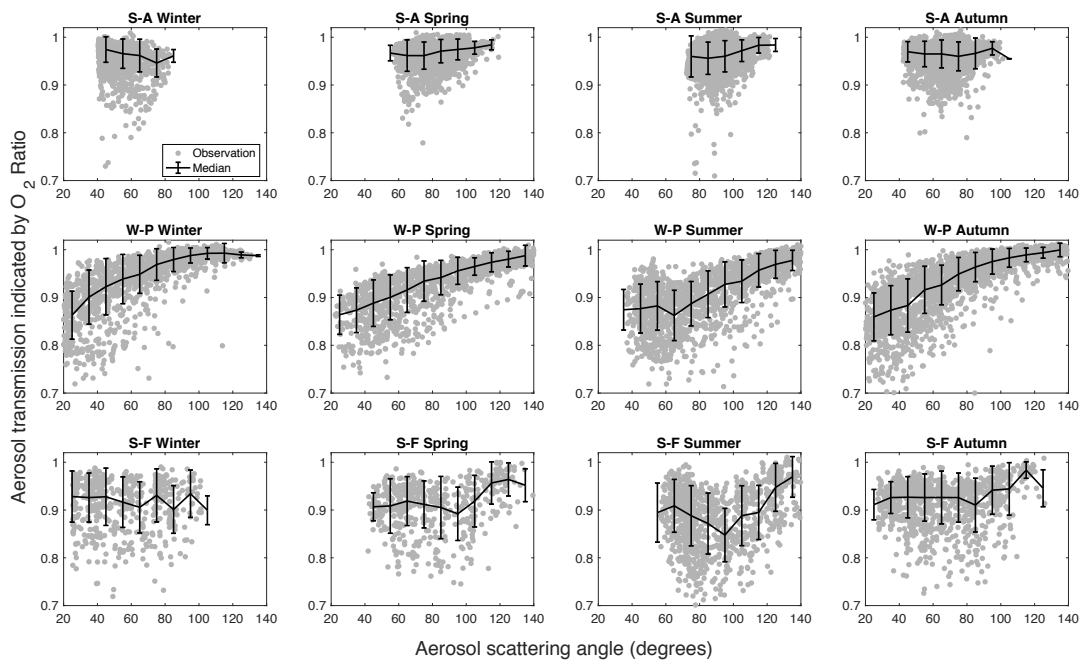

**Figure A4**. The same with **Figure 8** but with scattering angles as axis. Multiyear average of $O_2$ ratio for the three surface reflection points, Santa Anita (S-A) on the top, West Pasadena (W-P) in the middle, and Santa Fe Dam (S-F) at the bottom from CLARS-FTS data for all seasons (The original measurements in grey and the hourly mean in black). The corresponding simulation estimates using 2S-ESS RT model with averaged solar geometris, aerosol optical properties, and surface reflectance. All the data from 2011 to 2018 are used.